\documentclass{article}

\PassOptionsToPackage{round, compress}{natbib}

\usepackage[preprint]{neurips_2025}

\usepackage[utf8]{inputenc} 
\usepackage[T1]{fontenc}    
\usepackage{hyperref}       
\usepackage{url}            
\usepackage{booktabs}       
\usepackage{amsfonts}       
\usepackage{nicefrac}       
\usepackage{microtype}      
\usepackage{xcolor}         
\usepackage[pdftex]{graphicx}
\usepackage{amsmath}
\usepackage{multirow}

\makeatletter
\renewcommand{\@fnsymbol}[1]{\@arabic{#1}}
\makeatother

\title{Modeling Gene Expression Distributional Shifts for Unseen Genetic Perturbations
}

\author{%
  Kalyan Ramakrishnan \\
  University of Oxford \\
  \And
  Jonathan G. Hedley\thanks{\texttt{jonathan.hedley@eng.ox.ac.uk}} \\
  University of Oxford \\
  \And
  Sisi Qu \\
  University of Oxford \\
  \And
  Puneet K. Dokania \\
  University of Oxford \\
  \And
  Philip H. S. Torr \\
  University of Oxford\\
  \AND
  Cesar A. Prada-Medina \\
  Novo Nordisk\\
  \And
  Julien Fauqueur \\
  Novo Nordisk\\
  \And
  Kaspar M\"artens\thanks{\texttt{KQTM@novonordisk.com}} \\
  Novo Nordisk\\
}

\begin{document}

\maketitle

\begin{abstract}
We train a neural network to predict distributional responses in gene expression following genetic perturbations. This is an essential task in early-stage drug discovery, where such responses can offer insights into gene function and inform target identification. Existing methods only predict changes in the mean expression, overlooking stochasticity inherent in single-cell data. In contrast, we offer a more realistic view of cellular responses by modeling expression distributions. Our model predicts gene-level histograms conditioned on perturbations and outperforms baselines in capturing higher-order statistics, such as variance, skewness, and kurtosis, at a fraction of the training cost. To generalize to unseen perturbations, we incorporate prior knowledge via gene embeddings from large language models (LLMs). While modeling a richer output space, the method remains competitive in predicting mean expression changes. This work offers a practical step towards more expressive and biologically informative models of perturbation effects.
\end{abstract}

\section{Introduction}

Predicting changes in gene expression in response to genetic perturbations is a central goal in functional genomics and early-stage drug discovery. The aim is to understand how perturbing a gene (by switching it off, reducing its activity, or increasing its expression) affects the expression of other genes across the genome. Gene expression is stochastic -- even in genetically identical cells, levels fluctuate due to transcriptional noise, regulatory variation, and measurement uncertainty~\citep{PAULSSON2005, RAJ2008}. Thus, gene expression is best modeled as a distribution across cells, a crucial aspect many methods overlook.

\begin{figure}[t]
    \centering
    \includegraphics[width=0.84\linewidth]{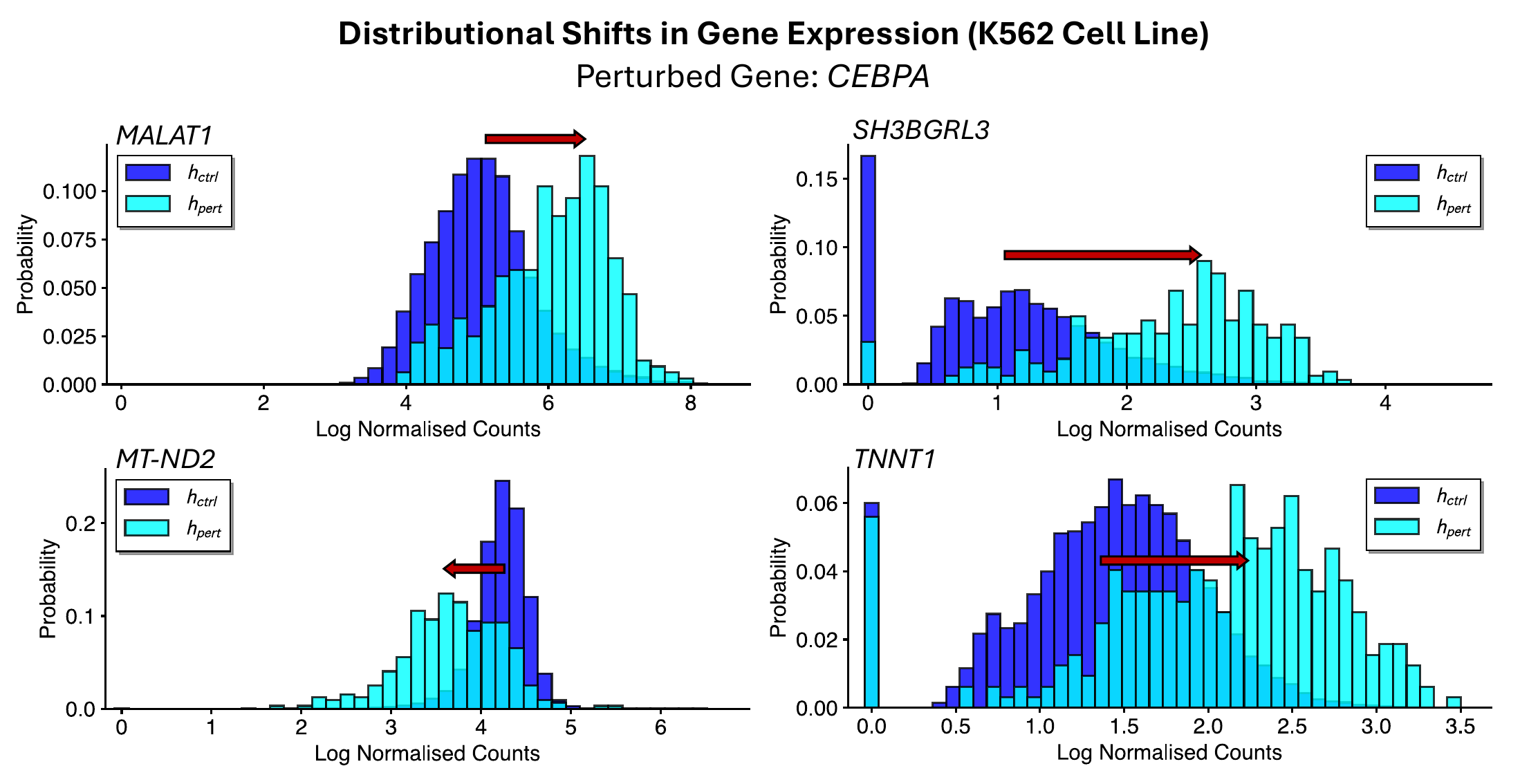}
    \caption{Gene expression distributional shifts following a CEBPA perturbation in the K562 cell line for four example genes. In each case, the control (dark blue) and perturbed (cyan) histograms differ in the mean \emph{and} overall shape, including variance, skewness, zero inflation, and multimodality.}
    \label{fig:dist}
\end{figure}

Evidence for the complexity of gene expression distributions comes from both observational and interventional data. For example, \citet{DETORRENTE2020} analyzed gene expression profiles from the Cancer Genome Atlas\footnote{\url{https://www.cancer.gov/tcga}} and cataloged a wide range of distribution shapes that more accurately reflect~biological variation. Incorporating these patterns into predictive models improved downstream tasks such as cancer survival prediction. Similar distributional diversity is seen in controlled perturbation experiments. Recent techniques such as Perturb-seq~\citep{DIXIT2016, NORMAN2019, REPLOGLE2022} enable expression profiling across thousands of single cells following genetic perturbations (gene knockouts), revealing shifts in mean expression, as well as in variance, skewness, and modality (Fig.~\ref{fig:dist}). However, experimentally profiling every possible perturbation in all cellular contexts is prohibitively expensive and time-consuming. This motivates the development of \emph{in silico} models that can predict responses to unseen perturbations, enabling generalization beyond the limited set of experimentally measured conditions.

Several deep generative models have been developed to predict gene expression responses to perturbations. CPA~\citep{LOTFOLLAHI2023} and sVAE+~\citep{LOPEZ2023} use latent-variable architectures with probabilistic decoders but rely on fixed parameterized likelihoods that may be too rigid to capture the diversity of expression patterns in real data~\citep{DETORRENTE2020}. More critically, they lack mechanisms for generalizing to unseen perturbations, limiting their use in exploratory perturbation response prediction~\citep{WU2024}. 

Other works focus on generalization by incorporating prior biological knowledge. GEARS~\citep{ROOHANI2024} uses gene ontology and co-expression graphs to propagate perturbation signals between related genes. Foundation models, such as scGPT~\citep{CUI2024} and scFoundation~\citep{HAO2024}, rely on transformers trained on large observational single-cell atlases to learn gene and cell embeddings. Despite their scale, these models only predict shifts in the mean and ignore distributional changes. Moreover, \citet{AHLMANN2025} show that these sophisticated models fail to outperform simple linear baselines when evaluated on out-of-distribution (OOD) perturbations. This suggests that scale and architectural complexity alone are insufficient for generalizable response prediction, motivating a shift toward simpler yet biologically grounded alternatives.

A promising direction is to embed prior knowledge directly into model inputs, enabling generalization via functional similarity. Large language model (LLM)-informed gene embeddings, derived from protein sequences, ontologies, or expression corpora, can capture functional relationships between genes~\citep{RIVES2021, THEODORIS2023}. By conditioning perturbation models on these embeddings, responses may be inferred for genes not seen during training. \citet{MARTENS2024} demonstrate that utilizing LLM-based embeddings significantly enhances OOD performance; however, their model only predicts mean expression changes. 

We address the above limitations by (i) predicting gene expression distributions and (ii) generalizing to unseen perturbations. To our knowledge, no existing method jointly tackles both. Our simple framework combines LLM-informed gene embeddings with a flexible histogram-based output, approximating per-gene expression distributions using discretized bins. On single-cell datasets~\citep{NORMAN2019, REPLOGLE2022}, this simple model outperforms other distributional baselines with as few as 15 bins and trains $\sim$3x faster. The richer output enables more informative analyses than mean shifts, such as identifying distinct responses to the same perturbation or estimating prediction uncertainty, helping better understand how genes influence one another.

\section{Related Work}

\textbf{Generative models for perturbation effects.} Generative models are widely used to simulate transcriptional responses in single cells. scVI~\citep{LOPEZ2018} models gene expression counts with a negative binomial VAE~\citep{kingma2013auto} to capture measurement noise and variability across cells. scGen~\citep{LOTFOLLAHI2019} and CPA~\citep{LOTFOLLAHI2023} learn latent-space shifts that encode perturbation effects, enabling interpolation across conditions. sVAE+~\citep{LOPEZ2023} and SAMS-VAE~\citep{BEREKET2023} separate baseline and perturbation components to improve interpretability, while CellOT~\citep{BUNNE2023} aligns control and perturbed populations via neural optimal transport. However, all remain limited to perturbations seen during training.

\textbf{Generalization to unseen perturbations.} Generalizing to unseen perturbations is critical for scalable profiling. GEARS~\citep{ROOHANI2024} addresses this by propagating perturbation signals over a gene ontology graph. More recently, foundation models such as scGPT~\citep{CUI2024} and scFoundation~\citep{HAO2024} learn gene and cell embeddings from large-scale data using transformers. Despite their scale, these models predict only mean responses and do not model distributions. Moreover, a recent benchmark~\citep{AHLMANN2025} shows they struggle to outperform simple baselines on OOD perturbations.

\textbf{Gene embeddings from language models.} Pretrained language models have recently been adopted to encode biological priors. GenePT~\citep{CHEN2024} generates embeddings from NCBI gene descriptions\footnote{\url{https://www.ncbi.nlm.nih.gov/gene}} and shows that these embeddings encode underlying biology, e.g., gene function. \citet{MARTENS2024} incorporate similar embeddings into perturbation models, demonstrating state-of-the-art generalization to unseen targets. However, all focus exclusively on mean responses. 

\section{Method}

\subsection{Overview}
We aim to model responses to genetic perturbations by learning a conditional distribution over gene expression values given a perturbation $p\in\{0,1\}^P$, where $P$ is the number of perturbable genes. Let $\mathbf{x}\in\mathbb{R}_{\geq 0}^{G}$ be a random vector representing post-perturbation log-normalized expression across $G$ genes. The task is to approximate the conditional distribution $q(\mathbf{x}|p)$. In principle, this is a joint distribution over all genes, residing in a high-dimensional space that is challenging to model directly (partly due to the sparsity and noise inherent in single-cell data). We therefore focus on predicting gene-wise marginal distributions $q(x_g|p)$ for each gene $g$. While this ignores gene-gene correlations, it allows tractable distribution modeling and captures per-gene variability across cells. 

\subsection{Histogram Construction}
To represent marginal expression distributions, we discretize each gene's expression range into a constant number of fixed-width bins and model the result as a histogram. Thus, we approximate $q(x_g|p)\approx h(x_g|p)$, where $h$ is a piecewise uniform density over the expression range. 

For each gene $g$, expression values in the training data (across all perturbations) are binned into $B$ intervals of width $w_g$. The test data is discretized using the same bin edges to prevent data leakage. Expression ranges are symmetrically extended by a small constant $\epsilon_g$ to avoid truncation:
\begin{equation}
    r_g =\left[\min(x_g)-\epsilon_g, \max(x_g)+\epsilon_g\right],\quad\mathrm{where}\quad \epsilon_g=\frac{\max(x_g)-\min(x_g)}{2(B-1)}
    \label{eq: rg}
\end{equation}
and the minimum and maximum values are calculated per gene from the training data. This choice of $\epsilon_g$ ensures the first bin is centered at the minimum, allowing the model to assign mass exactly at zero to better capture the sparsity in single-cell data. For genes with a constant expression (e.g., all zeros), we assign a small positive upper bound to avoid degenerate binning and ensure non-negative support. Histograms are then computed by assigning a probability mass to each bin based on the fraction of cells falling within it, yielding a valid discrete distribution.

\subsection{LLM-informed Perturbation Encoding}
We map each perturbation $p$ to a continuous vector $\mathbf{e}_{p} \in \mathbb{R}^{d}$ using a fixed function, where $d$ is the embedding dimension. For this, we first extract embeddings for each perturbed gene from the following two sources, building on the success of LLM-based models from~\citet{MARTENS2024}:
\begin{itemize}
    \item[(i)] \textbf{Gene descriptions:} Text embeddings of NCBI gene descriptions from GPT-3.5, as proposed by~\citet{CHEN2024}.
    \item[(ii)] \textbf{Protein sequences:} Sequence embeddings from the ProtT5~\citep{ELNAGGAR2022} protein language model.
\end{itemize}
We concatenate both types following~\citet{MARTENS2024}. These gene embeddings are frozen while training and serve as biologically informed inputs to enable generalization to unseen perturbations. If several genes are perturbed simultaneously, we could aggregate their embeddings via a simple average. However, we restrict ourselves to single-gene perturbations in this work.

\subsection{Model Architecture}
Given a perturbation, we model the induced \emph{distributional shift} in gene expression relative to an empirically observed control (unperturbed) distribution. This reflects the biological intuition that most perturbations modify, rather than overwrite, a cell's baseline expression profile. Let $h_{\rm ctrl}\in \mathbb{R}^{G\times B}$ denote the control histogram, where each row (summing to 1) contains the baseline distribution of a gene over the $B$ bins. For a perturbation $p$, we apply shifts $\Delta\in\mathbb{R}^{G\times B}$ in log space as follows:
\begin{equation}
    \hat{h}_\theta (p) = \mathrm{softmax} \left(\log h_{\rm ctrl}+\Delta_\theta(\mathbf{e}_{p}) \right),
\end{equation}
where $\Delta_\theta: \mathbb{R}^d \rightarrow \mathbb{R}^{G\times B}$ is a multilayer perceptron (MLP) with learnable parameters $\theta$ and output initialized close to zero~\citep{he2015delving}. The softmax is applied over each row. Note that $\Delta=0$ recovers the control histogram.

\subsection{Training Objective}
We train the model by minimizing a loss that compares the predicted ($\hat h$) and target ($h$) distributions using two components: (i) the Wasserstein-1 (W1) distance for distributional alignment and (ii) mean squared error (MSE) to match mean expressions. While cross-entropy is used for histogram prediction in other contexts~\citep{IMANI2024}, it treats bins independently and ignores the geometry of the support. In contrast, the W1 distance accounts for mass transport across bins and better reflects the continuity of expression data. 

\textbf{W1 term.} For general 1D distributions, the W1 distance is given by $\mathrm{W1} = \int_{-\infty}^{\infty}|\hat{F}(x)-F(x)|dx$, where $\hat{F}$ and $F$ are the predicted and true cumulative distribution functions (CDFs). For histogram distributions, we can compute this integral in closed form, as described below.

Let $I_{gb}=[x_{gb}^{(l)}, x_{gb}^{(r)}]$ be the $b$-th bin interval for gene $g$ (with width $w_g$), to which the predicted and true histograms assign masses $\hat{h}_{gb}$ and $h_{gb}$ given a perturbation. The CDF gaps at the left and right bin edges are
\begin{equation}
C_{gb}^{-} = \hat{F}_g(x_{gb}^{(l)})-F_g(x^{(l)}_{gb}) \quad\mathrm{and}\quad C_{gb}^{+} = \hat{F}_g(x_{gb}^{(r)})-F_g(x^{(r)}_{gb}).
\end{equation}
Assuming the densities are uniform inside $I_{gb}$, the gap varies linearly as follows (for $x\in I_{gb}$):
\begin{equation}
  \hat{F}_{g}(x)-F_{g}(x)
  =C_{gb}^{-}
   +\frac{\delta_{gb}}{w_{g}}\bigl(x-x_{gb}^{(l)}\bigr),
\label{eq:linear}
\end{equation}
where $\delta_{gb} = C_{gb}^{+} - C_{gb}^{-} = \hat{h}_{gb}-h_{gb}$ represents the overall change. Integrating Eq.~\ref{eq:linear} over $I_{gb}$ gives its exact contribution to the W1 distance:
\begin{equation}
    \mathrm{W1}_{gb}=\int_{I_{gb}}
              \lvert \hat{F}_g(x)-F_g(x)\rvert dx 
  =\begin{cases}
     \dfrac{w_{g}}{2}\bigl(\lvert C_{gb}^{-}\rvert
                               +\lvert C_{gb}^{+}\rvert\bigr)
       & \text{if}\ \ \ C_{gb}^{-}C_{gb}^{+}\ge 0\\[6pt]
     \dfrac{w_{g}}{2}\dfrac{\lvert C_{gb}^{-}\rvert^{2}
                               +\lvert C_{gb}^{+}\rvert^{2}}
                              {\lvert C_{gb}^{-}\rvert
                               +\lvert C_{gb}^{+}\rvert}
       &  \text{if}\ \ \ C_{gb}^{-}C_{gb}^{+}<0.
   \end{cases}
   \label{eq: W1}
\end{equation}
The W1 loss is thus obtained by summing Eq.~\ref{eq: W1} over the $B$ bins and averaging over all genes:
\begin{equation}
    \mathcal{L}_{\rm wass} = \frac{1}{G}\sum_{g=1}^{G}\sum_{b=1}^{B} \mathrm{W1}_{gb}.
\label{eq:lwass}
\end{equation}

\textbf{MSE term.} This term resolves cases where several histograms share the same W1 distance from the control distribution and is defined as
\begin{equation}
    \mathcal{L}_{\rm mse} = \frac{1}{G}\sum_{g=1}^{G}(\hat{\mu}_{g} - \mu_{g})^2,\quad \mathrm{where}\quad\hat{\mu}_{g} = \sum_{b=1}^{B} \frac{1}{2}(x_{gb}^{(l)} + x_{gb}^{(r)})\ \hat{h}_{gb}
\label{eq:lmse}
\end{equation}
is the predicted mean and $\mu_g$ is the target mean expression of gene $g$ (for the given perturbation).

\textbf{Total loss.} 
We average the two components over perturbations and combine them with fixed weights:
\begin{equation}
\mathcal{L}_{\rm total} = \lambda_{\rm wass}\mathcal{L}_{\rm wass} + \lambda_{\rm mse}\mathcal{L}_{\rm mse},
\end{equation}
where $\lambda_{\rm wass}+\lambda_{\rm mse}=1$. The loss $\mathcal{L}_{\rm total}$ is minimized over model parameters $\theta$. 

We provide Python code for our training procedure on GitHub\footnote{\url{https://github.com/Kalyan0821/LLMHistPert}}.

\section{Experiments}
\subsection{Experimental Details}
We evaluate our method on three single-cell Perturb-seq datasets across two cell lines. These include 102 single-gene perturbations in the K562 cell line from~\citet{NORMAN2019}, as well as 1076 (K562) and 1517 (RPE1) perturbations from~\citet{REPLOGLE2022}. Gene expression values are log-normalized. We train and evaluate all models on single-gene perturbations. We employ 9-fold cross-validation to evaluate generalization to unseen perturbations, given the limited dataset size. Training is done on a single NVIDIA A40 GPU (45 GB) and takes 10-15 minutes per fold, depending on histogram resolution. We use 15 bins by default.

Histogram models are trained with a weighted sum of the W1 and MSE losses. A coarse grid sweep identified the best configuration as $\lambda_{\rm wass} = 0.75$ and $\lambda_{\rm mse} = 0.25$ (see Appendix~\ref{app:hypers} for details). We vary the MLP size with dataset scale to reduce overfitting: hidden dimensions are $(1024, 512, 1024)$ for the Replogle data and $(128, 64)$ for the smaller Norman dataset. Hidden layers use Layer Normalization~\citep{ba2016layer} followed by a Leaky ReLU activation with negative slope $0.01$. The output layer is linear. Models are trained using the AdamW optimizer~\citep{loshchilov2017decoupled} with learning rate $10^{-3}$, weight decay $10^{-3}$, dropout $0.2$, and batch size $32$, for $500$ epochs per fold. Details of resource usage for training are provided in Appendix~\ref{app:resources}.

\subsection{Distributional Baselines \& Evaluation}
To assess how well our model captures post-perturbation gene expression distributions, we construct several baselines as described below.

First, \emph{Ctrl Hist} predicts the control distribution $h_{\rm ctrl}$ for all perturbations. This represents a naive baseline that assumes no response to intervention. Second, \emph{Non-Ctrl Hist} pools all perturbed cells in the training set and fits common per-gene histograms to the resulting data. This captures the aggregate expression distribution across perturbations -- which may fare well on average likelihood metrics -- but cannot model perturbation-specific effects.

Third, to create a perturbation-aware baseline, we train an MLP ($\mathbb{R}^d \rightarrow \mathbb{R}^{G\times 2}$) to predict the parameters of a truncated Gaussian (\emph{TG}) for each gene conditioned on the perturbation. This can capture coarse changes in the mean and variance, with truncation enforcing non-negative expressions and enabling fair comparison with the bounded supports used by the histogram model (Eq.~\ref{eq: rg}). The TG density has the following general form on its support $[a, b]$:
\begin{equation}
    f_{\rm TG}(x;\mu,\sigma) = \frac{\frac{1}{\sigma\sqrt{2\pi}}e^{-\frac{1}{2}\left(\frac{x-\mu}{\sigma}\right)^2}}{\Phi\big(\frac{b-\mu}{\sigma}\big)-\Phi\big(\frac{a-\mu}{\sigma}\big)}, \quad\mathrm{where}\quad\Phi(x)=\frac{1}{2}\left(1+\mathrm{erf}(x/\sqrt{2})\right)
\end{equation}
and parameters $(\mu, \sigma)$ describe the underlying Gaussian truncated to $[a, b]$. In our case, given perturbation $p$, the model outputs $G$ pairs $\left(\mu_{g,\theta}(p),\ \log\sigma_{g,\theta}(p)\right)$ corresponding to supports $r_g=[a_g, b_g]$. We also constrain $\mu_{g,\theta}$ to $[a_g-1.5\sigma_U, b_g+1.5\sigma_U]$, where $\sigma_U=(b_g-a_g)/\sqrt{12}$ is the standard deviation of a uniform distribution over $r_g$. This prevents the model from pushing $\mu_{g,\theta}$ far outside the observed data range, which may cause vanishing likelihoods and unstable training.

Finally, to account for the prevalence of zero-valued observations in single-cell data, we define a zero-inflated truncated Gaussian (\emph{ZITG}) density with the following general form:
\begin{equation}
        f_{\rm ZITG}(x; \mu, \sigma, \pi)=\pi~\mathcal{U}_{[-\epsilon,\epsilon]}(x)+(1-\pi)~f_{\rm TG}(x;\mu,\sigma),
\end{equation}
where the extra parameter $\pi \in [0, 1]$ represents the zero-inflation probability and $\mathcal{U}_{[-\epsilon,\epsilon]}$ is a narrow uniform density of width $2\epsilon$ centred at the origin. In our case, given perturbation $p$, the MLP ($\mathbb{R}^d \rightarrow \mathbb{R}^{G\times 2}\times[0, 1]^G$) produces $G$ triplets $\left(\mu_{g,\theta}(p),\ \log\sigma_{g,\theta}(p),\ \pi_{g,\theta}(p)\right)$ corresponding to the supports $r_g$ and half-widths $\epsilon_g$ defined in Eq.~\ref{eq: rg}.

\textbf{Baseline training.} For the TG and ZITG baselines, we minimize the average negative log-likelihood (NLL) of the training data. Given a perturbation, the NLL of gene $g$'s expressions is
\begin{equation}
    \mathrm{NLL}_g = -\frac{1}{N_g}\sum_{n=1}^{N_g}\log{f(x_{gn}; \theta}),
\end{equation}
where $N_g$ is the number of post-perturbation samples of gene $g$. The average NLL is then obtained by averaging this over all genes and perturbations in the dataset. 

\textbf{Evaluating distributions.} For each method, we report the average NLL over unseen perturbations to test if the predicted distributions fit the observed expression values. Note that for the histogram model, NLLs can be obtained efficiently without evaluating on the samples $\{x_{gn}\}$ each time as follows:
\begin{equation}
    \mathrm{NLL}^{\mathrm{hist}}_g = -\sum_{b=1}^{B}h_{gb}\log \left(\frac{\hat{h}_{gb}}{w_g}\right).
\end{equation}

We also test the ability to represent perturbation-specific structure by reporting errors in \emph{statistics} of the predicted distributions compared to the observed data. Specifically, we compute the relative mean absolute errors (RMAE) of the predicted mean, variance, skewness, and excess kurtosis. Statistics are obtained from the parameters of the TG / ZITG and the bin probabilities of the histogram model (see Appendix~\ref{app:stats} for details). Given an unseen perturbation, the RMAE of statistic $\tau$ is given by
\begin{equation}
    \mathrm{RMAE}(\hat\tau, \tau) = \dfrac{\sum_{g=1}^{G}|\hat{\tau}_{g}-\tau_{g}|}{\sum_{g=1}^{G}|\tau_{g}|},
\end{equation}
where $\hat{\tau}_{g}$ and $\tau_{g}$ denote its predicted and observed value for gene $g$. We average this error over all perturbations in the test set, serving as a ``shape-aware'' distributional metric.

\begin{table}[t]
    \centering
    \begin{tabular}{c c c c c c c c}
    \toprule
    \multirow{2}{*}{\textbf{Cell line}} & \multirow{2}{*}{\textbf{Method}} & \multirow{2}{*}{\textbf{NLL} ($\downarrow$)} & \multicolumn{4}{c}{\textbf{RMAE} ($\downarrow$)}                            & \textbf{Train time} ($\downarrow$) \\
                                        &                                 &                                              & \textbf{Mean} & \textbf{Var} & \textbf{Skew} & \textbf{Kurt}                & (mins/fold) \\ 
    \midrule
    \multirow{5}{*}{K562}
    & Ctrl Hist   & $-$0.379 & 1.000 & 0.184 & 0.237 & 0.572 & N/A\\
    & Non-Ctrl Hist & \bfseries $-$0.393 & 0.980 & 0.213 & 0.235 & 0.590 & N/A\\
    & TG (LLM)        & \ \ \ 0.016 &  0.934 & 0.250 & 0.472 &  0.940 & 20.49\\
    & ZITG (LLM)                         & $-$0.381 &  0.895 &  0.182 & 0.241 &   0.600 & 24.52\\
    & MLP+Hist (LLM)                     & $-$0.388 & \bfseries 0.879  & \bfseries 0.175  & \bfseries 0.225  &  \bfseries 0.539 & \bfseries 9.89\\
    \midrule
    \multirow{5}{*}{RPE1}
    & Ctrl Hist         & $-$0.437 & 1.000 & 0.269 & 0.308 & 0.685 & N/A\\
    & Non-Ctrl Hist     & \bfseries $-$0.459 & 0.895 & 0.266 & 0.284 & 0.665 & N/A\\    
    & TG (LLM)                & $-$0.068 & 0.896 & 0.302 & 0.461 & 0.899 & 41.16\\
    & ZITG (LLM)             & $-$0.455 & 0.865 & 0.242 & 0.281 & 0.655 & 49.49\\
    & MLP+Hist (LLM)  & $-$0.441 & \bfseries 0.863 & \bfseries 0.239 & \bfseries 0.276 & \bfseries 0.633 & \bfseries 14.93 \\
    \bottomrule
    \end{tabular}\vspace{0.3cm}
    \caption{Distributional model performance on unseen perturbations from the two Replogle cell lines. For each method, we report the gene-averaged NLL and RMAEs of post-perturbation statistics.}
    \label{tab:quant}
\end{table}

\begin{figure}[t]
    \centering
    \includegraphics[width=0.83\linewidth]{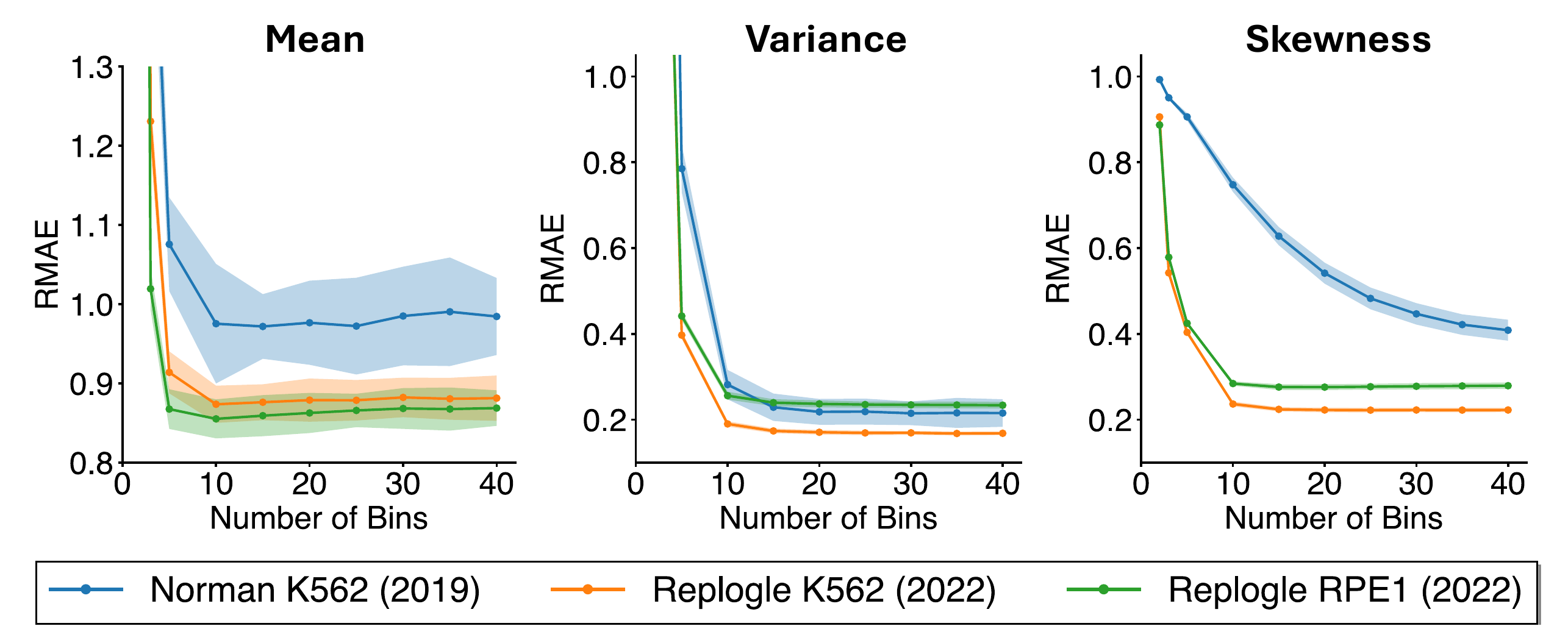}
    \caption{Effect of histogram resolution on the 
    prediction error of post-perturbation statistics
    for three benchmark datasets. Performance improves with resolution but plateaus at 15-20 bins on the larger Replogle datasets. Shading indicates the standard deviation across folds.}
    \label{fig:res_dist}
\end{figure}

\subsection{Quantitative Results}
Table~\ref{tab:quant} compares our histogram model (\emph{MLP+Hist}) to the distributional baselines on unseen perturbations from the two Replogle cell lines. On both, \emph{MLP+Hist} achieves the lowest RMAE for all four statistics, with the improvement over \emph{ZITG} increasing with the moment order (e.g., from 1.8\% in the mean to 10.2\% in kurtosis, on K562). This behavior is learned entirely from the expression data, unlike \emph{TG} / \emph{ZITG} that fix the functional form of the distribution, enabling greater flexibility to represent perturbation-specific structure. As expected, \emph{ZITG} outperforms \emph{TG} on all metrics, given the extra flexibility from the zero-inflation component.

Moreover, Table~\ref{tab:quant} shows that \emph{MLP+Hist} offers \textbf{significantly higher data efficiency}, training $\sim$2.5x faster than \emph{ZITG} on K562 and 3.3x faster on RPE1. We also analyze the memory usage of the methods in Appendix~\ref{app:resources}, where \emph{MLP+Hist} has only a fraction of the requirements of \emph{TG} / \emph{ZITG}. This is because rather than optimizing for hundreds of post-perturbation samples per gene, \emph{MLP+Hist} trains on a fixed-size, compressed version of this data (15 bin heights per gene). Still, our method generalizes more effectively to unseen perturbations, demonstrating that learning histogram representations is a straightforward and scalable approach to modeling genetic perturbation effects.

Interestingly, \emph{Non-Ctrl Hist} attains the best NLL on both cell lines. This can be attributed to the pooled (global) histogram assigning a substantial mass to the zero-expression bin ($\sim$61\% in K562 and 64\% in RPE1), which accounts for the majority of single-cell measurement data. However, since it predicts the same histogram for every perturbation, it cannot represent perturbation-specific shifts and performs significantly worse on the shape-aware metrics (Table~\ref{tab:quant}).

\begin{figure}
    \centering
    \includegraphics[width=0.75\linewidth]{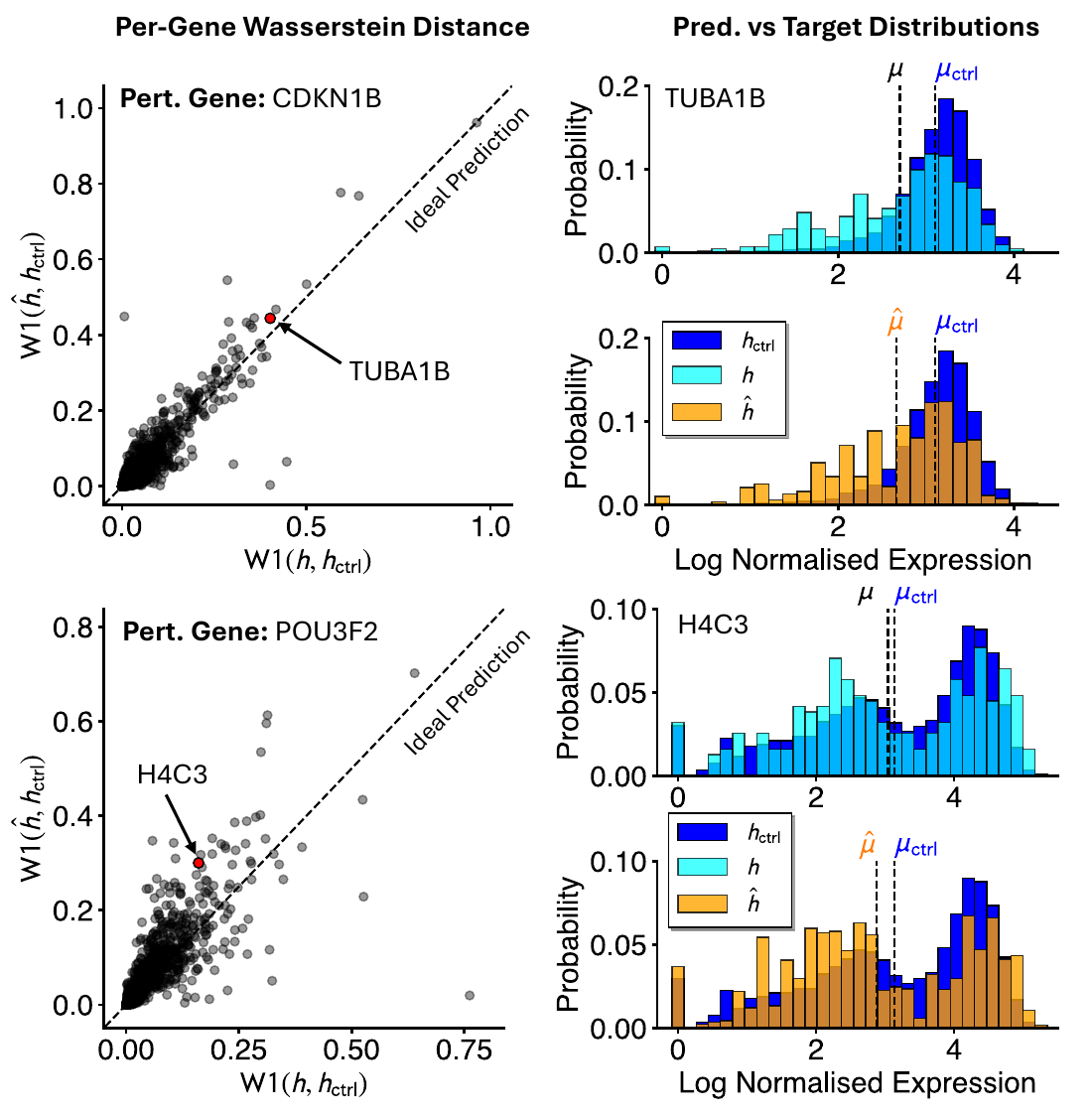}
    \caption{Distributional shift plots (left) and highlighted gene expression histograms (right) (using 35 bins) following two example perturbations (CDKN1B and POU3F2) from the Norman data. Each circle represents a gene $g$ with coordinates $x = \mathrm{W1}(h_{g}, h^{\mathrm{ctrl}}_g)$ and $y = \mathrm{W1}(\hat{h}_{g}, h^{\mathrm{ctrl}}_g))$, measuring the observed and predicted shifts from the control distribution (dark blue), respectively. We include histograms for two genes that exhibit non-mean effects: (i) TUBA1B, which shows increased variance, and (ii) H4C3, which has a multimodal response. In both cases, predicted histograms (orange) closely match the post-perturbation data (cyan). Vertical dashed lines mark the means.}
    \label{fig:wass_plot}
\end{figure}

\textbf{Effect of Histogram Resolution.}
Fig.~\ref{fig:res_dist} shows how model performance varies with histogram resolution. On the Replogle data, as the number of bins increases, the RMAEs of the mean, variance, and skewness steadily improve before plateauing at 15–20 bins, suggesting that a moderate resolution is sufficient to recover statistical features while maintaining computational efficiency. The improvement is less consistent across statistics on the Norman dataset, likely due to its relatively small training set size of around 90 perturbations per fold, compared to 1000 in Replogle.

\subsection{Qualitative Results}
To qualitatively assess the predicted distributional shifts, we plot W1 distances between the control and predicted distributions $\mathrm{W1}(\hat{h}_{g}, h^{\mathrm{ctrl}}_g)$ vs.~those between the control and observed distributions $\mathrm{W1}(h_{g}, h^{\mathrm{ctrl}}_g)$ for all genes $g$ given an unseen perturbation. Fig.~\ref{fig:wass_plot} (left) shows this plot for two example perturbations (CDKN1B\footnote{\url{https://www.ncbi.nlm.nih.gov/gene/1027}} and POU3F2\footnote{\url{https://www.ncbi.nlm.nih.gov/gene/5454}}), where each circle represents a gene. The plot provides a global view of distributional shift accuracy, helping identify genes where predictions succeed or fail (ideal predictions lie along the diagonal). We examine two genes (TUBA1B\footnote{\url{https://www.ncbi.nlm.nih.gov/gene/10376}} and H4C3\footnote{\url{https://www.ncbi.nlm.nih.gov/gene/8364}}) exhibiting distinct non-mean effects that our model recovers effectively.

TUBA1B encodes a tubulin alpha chain involved in building the cell's internal structure and regulating cell-cycle progression~\citep{WANG2024}. Its expression in control cells is tightly clustered. CDKN1B (perturbed gene) restrains cell division by putting a checkpoint on the DNA replication phase -- when this gene is knocked down, the checkpoint is lost, causing cells to progress through the cycle asynchronously~\citep{SUN2016}. As a result, the TUBA1B expression becomes more variable across the population. As seen in Fig.~\ref{fig:wass_plot} (top right), \emph{MLP+Hist} captures this increase in variance (and reduction in mean), demonstrating the ability to predict higher-order effects beyond the mean.

H4C3, a histone gene involved in DNA packaging and gene regulation~\citep{SEAL2022}, shows a bimodal distribution in control conditions, indicating distinct transcriptional states. When POU3F2 is knocked down, more cells enter the low-expression state, and fewer remain high~\citep{EISEN1995}. As seen in Fig.~\ref{fig:wass_plot} (bottom right), \emph{MLP+Hist} captures this change in proportions (although the mean hardly changes and peak positions remain stable), highlighting the ability to predict subtle yet biologically relevant shifts.

Detecting variance inflation in a cell-cycle gene or shifts in chromatin states provides mechanistic insights inaccessible to point-estimate methods. Thus, the histogram model has the potential to offer a more informative basis for biological hypothesis generation than models limited to mean expression.

\begin{table}[t]
    \centering
    \begin{tabular}{c c c c c c}
    \toprule
    \multirow{2}{*}{\textbf{Cell line}} & \multirow{2}{*}{\textbf{Method}} & \multicolumn{3}{c}{\textbf{Pearson correlation} ($\uparrow$)} & \multirow{2}{*}{\textbf{RMAE} ($\downarrow$)} \\
                                        &                                 & \textbf{Top 20} & \textbf{Top 50} & \textbf{Top 100}          & \\
    \midrule
    \multirow{5}{*}{K562}
    & Non-Ctrl Mean     & 0.460 \scriptsize{$\pm$ 0.03} & 0.495 \scriptsize{$\pm$ 0.02} & 0.513 \scriptsize{$\pm$ 0.02} & 0.981 \scriptsize{$\pm$ 0.01} \\
    & GEARS           & 0.472 \scriptsize{$\pm$ 0.03} & 0.504 \scriptsize{$\pm$ 0.03} & 0.513 \scriptsize{$\pm$ 0.03} & 1.050 \scriptsize{$\pm$ 0.02} \\
    & GP (LLM)        & 0.656 \scriptsize{$\pm$ 0.03} & \bfseries 0.702 \scriptsize{$\pm$ 0.03} & 0.706 \scriptsize{$\pm$ 0.02} & \bfseries 0.869 \scriptsize{$\pm$ 0.02} \\
    & MLP+Mean (LLM)                    & \bfseries 0.658 \scriptsize{$\pm$ 0.02} & 0.701 \scriptsize{$\pm$ 0.02} & \bfseries 0.710 \scriptsize{$\pm$ 0.02} & 0.871 \scriptsize{$\pm$ 0.03} \\
    & MLP+Hist (LLM)               & 0.653 \scriptsize{$\pm$ 0.03} & 0.697 \scriptsize{$\pm$ 0.02} & 0.703 \scriptsize{$\pm$ 0.02} & 0.876 \scriptsize{$\pm$ 0.02} \\
    \midrule
    \multirow{4}{*}{RPE1}
    & Non-Ctrl Mean& 0.654 \scriptsize{$\pm$ 0.03} & 0.702 \scriptsize{$\pm$ 0.02} & 0.723 \scriptsize{$\pm$ 0.02} & 0.893 \scriptsize{$\pm$ 0.03} \\
    & GEARS           & 0.553 \scriptsize{$\pm$ 0.02} & 0.621 \scriptsize{$\pm$ 0.03} & 0.660 \scriptsize{$\pm$ 0.03} & 0.955 \scriptsize{$\pm$ 0.03} \\
    & MLP+Mean (LLM)                    & \bfseries 0.685 \scriptsize{$\pm$ 0.03} & \bfseries 0.733 \scriptsize{$\pm$ 0.02} & \bfseries 0.756 \scriptsize{$\pm$ 0.02} & 0.860 \scriptsize{$\pm$ 0.03} \\
    & MLP+Hist (LLM)               & 0.672 \scriptsize{$\pm$ 0.03} & 0.726 \scriptsize{$\pm$ 0.02} & 0.749 \scriptsize{$\pm$ 0.02} & \bfseries 0.859 \scriptsize{$\pm$ 0.02} \\
    \bottomrule
    \end{tabular}\vspace{0.3cm}
    \caption{Mean prediction performance ($\pm$ standard deviation across folds) on unseen perturbations from the two Replogle cell lines. For each method, we report correlations (between predicted and observed mean shifts) over the top differentially expressed (DE) genes and RMAE over all genes.}
    \label{tab:quant_mean}
\end{table}

\begin{figure}[t]
    \centering
    \includegraphics[width=0.83\linewidth]{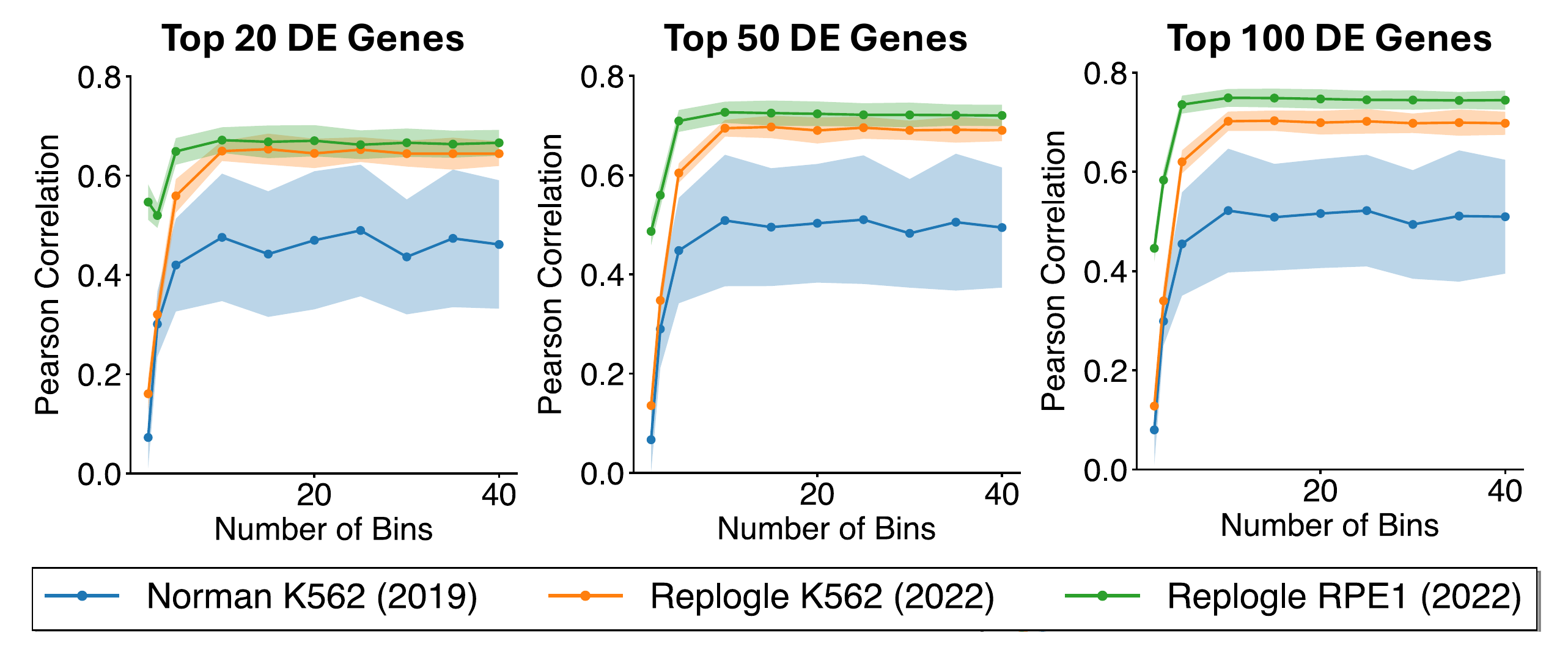}
    \caption{Effect of histogram resolution on the correlation between predicted and true mean shifts over the top differentially expressed (DE) genes for three benchmark datasets. Performance improves with resolution but plateaus at 15-20 bins. Shading indicates the standard deviation across folds.
    }
    \label{fig:res_mean}
\end{figure}

\subsection{Comparison to Mean-based Methods}
We evaluate mean prediction performance separately to compare \emph{MLP+Hist} with existing methods that predict only mean expression shifts (as a sanity check). Of these, we consider the graph-based \emph{GEARS}~\citep{ROOHANI2024} and the recent LLM-based Gaussian process regressor (\emph{GP}) by \citet{MARTENS2024}. We also consider two baselines: (i) \emph{Non-Ctrl Mean}, which predicts the per-gene mean expression over all perturbed cells in the training set -- incorporating global response trends but not perturbation identity, and (ii) \emph{MLP+Mean}, where we train an MLP ($\mathbb{R}^d\to\mathbb{R}^G$) with the same architecture as \emph{MLP+Hist} to predict per-gene shifts in mean expression -- representing a perturbation-aware baseline. We exclude methods like scGPT~\citep{CUI2024} since it was shown to be easily outperformed by \emph{GEARS} / \emph{Non-Ctrl Mean}~\citep{AHLMANN2025, MARTENS2025}.

Following prior work~\citep{BEREKET2023, ROOHANI2024}, for each method, we report the perturbation-averaged (i) Pearson correlation between predicted ($\mathbf{\hat{\mu}}-\mathbf{\mu_{\rm ctrl}}$) and observed ($\mathbf{\mu}-\mathbf{\mu_{\rm ctrl}}$) mean shifts -- calculated separately over the top 20, 50, and 100 differentially expressed (DE) genes, and (ii) RMAE of the predicted mean shift. Focusing on differences relative to control penalizes models that replicate the mean control expression, which can have a deceptively high correlation with the observed mean~\citep{MARTENS2024}.

Table~\ref{tab:quant_mean} compares the mean-based methods on unseen perturbations from the two Replogle cell lines. On both, \emph{MLP+Hist} performs similarly to the other LLM-based models, including the dedicated mean regressor \emph{MLP+Mean}, while modeling a richer output space. It also outperforms \emph{Non-Ctrl Mean} and earlier approaches like GEARS on all metrics. Interestingly, the simple \emph{Non-Ctrl Mean} baseline significantly outperforms GEARS on RPE1 and gets close on K562, further highlighting the advantage of LLM-derived biological priors for generalizable perturbation response prediction.

\textbf{Effect of Histogram Resolution.}
Fig.~\ref{fig:res_mean} shows how \emph{MLP+Hist} performance varies with histogram resolution. Correlations improve with increasing bin count and plateau at 15-20 bins, confirming that a moderate resolution is sufficient to recover shifts in mean expression as well.

\section{Conclusion}
We introduced a neural network that predicts gene expression distributional shifts using LLM embeddings of unseen genetic perturbations with a histogram-based output. Through experiments on standard Perturb-seq datasets, we showed that this simple method (i) outperforms baselines in predicting distributional structure at a much lower training cost, (ii) can reveal mechanistic insights that may enable improved hypothesis generation in genomics, and (iii) is competitive with other methods in predicting mean expression changes. It also highlights the benefit of LLM-based biological priors for generalization to unseen perturbations.

We end with some limitations and suggestions for future work. First, predictions are constrained by the expression range observed during training due to the histogram representation, which limits extrapolation and requires truncated baselines for fair comparison. Second, the method treats genes independently by modeling marginal distributions but not the regulatory dependencies between genes. Future work could explore methods for efficiently modeling this high-dimensional joint distribution. Third, while we use a simple discrete representation, continuous alternatives, such as kernel density estimation (KDE), could improve flexibility and extrapolation, albeit with higher modeling complexity. Finally, future work should focus more on extracting valuable biological insights from distributional models to assess the practical utility of our approach better.

\section*{Acknowledgements}
KR is supported by the EPSRC Centre for Doctoral Training in Autonomous Intelligent Machines and Systems EP/S024050/1. JGH is supported by a Novo Nordisk Postdoctoral Fellowship in partnership with the University of Oxford. SQ is supported by Wellcome Trust in partnership with the University of Oxford. 

\bibliographystyle{abbrvnat}
\bibliography{neurips_2025}

\newpage

\appendix

\Large{\textbf{Appendix}}
\normalsize
\section{Higher-order Statistics}
\label{app:stats}
We calculate the variance, skewness, and excess kurtosis of a distribution from its raw moments $m_k = \mathbb{E}[X^k]$, where $k\in\{1, 2, 3, 4\}$, using the following formulas:
\[
\begin{aligned}
  \operatorname{Var}[X]
  &=m_2-m_1^2 \\
  \operatorname{Skew}[X]
  &=\frac{m_3 - 3\ m_1 m_2 + 2\ m_1^3}
         {\operatorname{Var}[X]^{3/2}} \\
  \operatorname{Kurt}[X]         
  &=\frac{m_4 - 4\ m_1 m_3 + 6\ m_1^2 m_2 - 3 m_1^4}
         {\operatorname{Var}[X]^{2}} - 3.
\end{aligned}
\]
We specify the moment expressions for each distribution below.

\textbf{Histogram.} Given a histogram with $B$ bin intervals $[x_{b}^{(l)}, x_{b}^{(r)}]$ of widths $w_b$ having probabilities $h_b$,
\[
  m_k^{\mathrm{hist}} = \sum_{b=1}^{B} \frac{(x_{b}^{(r)})^{k+1} - (x_{b}^{(l)})^{k+1}}{k + 1} \frac{h_{b}}{w_{b}}.
\]

\textbf{TG.} Given parameters $(\mu, \sigma)$ and support $[a, b]$, 
\[
  m_k^{\mathrm{TG}} = \sum_{r=0}^{k} \binom{k}{r}\ \mu^{k-r} \sigma^r L_r,
\]
with
\[
\begin{aligned}
  L_0 &= 1 \\
  L_{1} &= -\ \frac{\varphi(\beta)-\varphi(\alpha)}{\Phi(\beta)-\Phi(\alpha)} \\
  L_{r} &= -\ \frac{\beta^{r-1}\varphi(\beta)-\alpha^{r-1}\varphi(\alpha)}{\Phi(\beta)-\Phi(\alpha)} + (r-1)\ L_{r-2}, \quad r\ge2,
\end{aligned}
\]
where $\alpha=(a-\mu)/\sigma$, $\beta=(b-\mu)/\sigma$, and $\varphi(\cdot)$ and $\Phi(\cdot)$ are the density and CDF of the standard normal distribution\footnote{\url{https://people.sc.fsu.edu/~jburkardt/presentations/truncated_normal.pdf}}.

\textbf{ZITG.} Given zero-inflation probability $\pi$, TG parameters $(\mu, \sigma)$, and support $[a, b]$,
\[
  m_k^{\mathrm{ZITG}} = \pi~m_k^{\mathcal{U}} + (1-\pi)~m_k^{\mathrm{TG}},
\]
where the moments of the uniform spike $\mathcal{U}_{[-\epsilon,\epsilon]}$ are given by
\[
  m_k^{\mathcal{U}} = \frac{1 - (-1)^{k+1}}{k + 1} \frac{\epsilon^{k}}{2}.
\]

\section{Resource Use for Training}
\label{app:resources}
Table~\ref{tab:resource} compares the training time, peak GPU memory usage, and learnable parameters of several methods described in this work for the two Replogle cell lines. Among the learnable distributional models, \emph{MLP+Hist} is both the quickest to train and the lightest in terms of memory usage despite carrying more parameters. For example, on the RPE1 data, it trains $\sim$2.7 - 3.3x faster than the \emph{TG} / \emph{ZITG} baselines and requires $\sim$8.2x less memory to train. 

The extra cost for the distributional baselines is due to the need to calculate and optimize log-likelihoods for every post-perturbation sample per gene. Moreover, since the number of cells varies across perturbations, the data is padded up to the largest sample size in its mini-batch, wasting computation and increasing memory usage. In contrast, \emph{MLP+Hist} learns from a fixed-size compression of the data, enabling simple and scalable modeling of genetic perturbation effects.

\begin{table}[t]
  \centering
    \begin{tabular}{c c c c c}
    \toprule
    \multirow{2}{*}{\textbf{Cell line}} & \multirow{2}{*}{\textbf{Method}}  & \textbf{Train time} & \textbf{GPU Memory Usage} & \multirow{2}{*}{\textbf{Learnable Params}} \\
                                        &                                   & (mins per fold)     & (MB)                      & \\
    \midrule
    \multirow{6}{*}{K562}
    & Non-Ctrl Mean   & N/A & 1,013 & N/A \\
    & Non-Ctrl Hist   & N/A & 1,617 & N/A \\
    & MLP+Mean (LLM)  & 3.14 & 1,575 & 8,802,696 \\
    & MLP+Hist (LLM)  & 9.89 & 3,219 & 80,552,696 \\
    & TG (LLM)        & 20.49 & 10,735 & 13,927,696 \\
    & ZITG (LLM)      & 24.52 & 12,023 & 19,052,696 \\
    \midrule
    \multirow{6}{*}{RPE1}
    & Non-Ctrl Mean   & N/A & 1,091 & N/A \\
    & Non-Ctrl Hist   & N/A & 1,885 & N/A \\
    & MLP+Mean (LLM)  & 4.47 & 1,655 & 8,802,696 \\
    & MLP+Hist (LLM)  & 14.93 & 3,503 & 80,552,696 \\
    & TG (LLM)        & 41.16 & 28,970 & 13,927,696 \\
    & ZITG (LLM)      & 49.49 & 28,710 & 19,052,696 \\
    \bottomrule
    \end{tabular}\vspace{0.3cm}
  \caption{Training costs of methods when trained on a single NVIDIA A40 GPU.}
  \label{tab:resource}
\end{table}

\section{Hyperparameters}
\label{app:hypers}
We experiment with using a weighted sum of three loss terms,
\[
\mathcal{L}_{\rm total} = \lambda_{\rm ce}\mathcal{L}_{\rm ce} + \lambda_{\rm wass}\mathcal{L}_{\rm wass} + \lambda_{\rm mse}\mathcal{L}_{\rm mse},
\]
where $\lambda_{\rm ce}+\lambda_{\rm wass}+\lambda_{\rm mse}=1$ and $\mathcal{L}_{\rm ce}$ is a simple cross-entropy loss (which we average over perturbations in the training set):
\[
\mathcal{L}_{\rm ce} = -\frac{1}{G}\sum_{g=1}^{G} \sum_{b=1}^{B}h_{gb}\log \hat{h}_{gb}.
\]
We perform a coarse grid search (with step size $0.25$) over valid triples $(\lambda_{\rm ce}, \lambda_{\rm wass}, \lambda_{\rm mse})$ and report mean-shift prediction performance for the Replogle K562 cell line in Table~\ref{tab:hist_loss_weights}. The best configuration ($\lambda_{\text{wass}} = 0.75$ and $\lambda_{\text{mse}} = 0.25$) assigns a majority weight to the Wasserstein term and retains the MSE contribution. Adding even a small weight to the cross-entropy loss hurts performance, with the worst configuration assigning all of the weight to it.

\begin{table}[h]
  \centering
    \begin{tabular}{ccc|cc}
    \toprule
        $\lambda_{\rm ce}$ & $\lambda_{\rm wass}$ & $\lambda_{\rm mse}$ & \textbf{Corr. Top 20} ($\uparrow$)\\
    \midrule
        1 & 0 & 0 & 0.628 \\ 
        0.5 & 0 & 0.5 & 0.635 \\
        0.25 & 0 & 0.75 & 0.636 \\
        0.75 & 0 & 0.25 & 0.638 \\        
        0.75 & 0.25 & 0 & 0.639 \\         
        0.25 & 0.5 & 0.25 & 0.639 \\
        0.25 & 0.75 & 0 & 0.644 \\        
        0.25 & 0.25 & 0.5 & 0.644 \\
        0.5 & 0.5 & 0 & 0.647 \\ 
        0 & 1 & 0 & 0.647 \\
        0 & 0.25 & 0.75 & 0.647 \\
        0 & 0.5 & 0.5 & 0.648 \\
        \textbf{0} & \textbf{0.75} & \textbf{0.25} & \textbf{0.651} \\        
    \bottomrule
    \end{tabular}\vspace{0.3cm}
  \caption{Effect of the loss weights used to train the histogram model.}
  \label{tab:hist_loss_weights}
\end{table}

\end{document}